# PSEUDOGAP EFFECTS ON THE CHARGE DYNAMICS IN THE UNDERDOPED COPPER OXIDE MATERIALS

SHIPING FENG, FENG YUAN
*Department of Physics, Beijing Normal University, Beijing 100875, China*

Within the $t$-$J$ model, the charge dynamics of copper oxide materials in the underdoped regime is studied based on the fermion-spin theory. It is shown that both in-plane charge dynamics and c-axis charge dynamics are mainly governed by the scattering from the in-plane fluctuation, which would be suppressed when the holon pseudogap opens at low temperatures, leading to the temperature linear to the non-linear range in the in-plane resistivity and crossovers to the semiconducting-like range in the c-axis resistivity.

It has become clear in the past several years that copper oxide materials are among the most complex systems studied in condensed matter physics, and show many unusual normal-state properties [1]. Among the striking features of the unusual normal-state properties stands out the extraordinary charge dynamics. The charge response, as manifested by the optical conductivity and resistivity, has been extensively studied experimentally [1,2] in the underdoped copper oxide materials, and the results indicate that (1) the in-plane conductivity shows the non-Drude behavior at low energies and anomalous midinfrared band in the charge-transfer gap, while the in-plane resistivity $\rho_{ab}(T)$ exhibits a temperature linear dependence with deviations at low temperatures; (2) for the copper oxide materials without the Cu-O chains in between the $CuO_2$ planes, the transferred weight in the c-axis conductivity forms a band peaked at high energy $\omega \sim 2ev$, and the low-energy spectral weight is quite small and spread over a wide energy range instead of forming a peak at low energies, the behavior of the c-axis temperature dependent resistivity $\rho_c(T)$ is characterized by a crossover from the high temperature metallic-like to the low temperature semiconducting-like; (3) for these copper oxide materials with the Cu-O chains in between the $CuO_2$ planes, the c-axis conductivity exhibits the non-Drude behavior at low energies and weak midinfrared band, but the c-axis resistivity $\rho_c(T)$ shows a crossover from the high temperature metallic-like behavior to the low temperature semiconducting-like behavior.

Among the microscopic models the most helpful for the discussion of the physical properties of the doped antiferromagnet is the $t$-$J$ model [3],

$$H = -t \sum_{\langle ij \rangle \sigma} C_{i\sigma}^{\dagger} C_{j\sigma} + h.c. - \mu \sum_{i\sigma} C_{i\sigma}^{\dagger} C_{i\sigma} + J \sum_{\langle ij \rangle} \mathbf{S}_i \cdot \mathbf{S}_j, \quad (1)$$

supplemented by the on-site local constraint $\sum_{\sigma} C_{i\sigma}^{\dagger} C_{i\sigma} \leq 1$ to avoid the double occupancy, where $C_{i\sigma}^{\dagger}$ ($C_{i\sigma}$) are the electron creation (annihilation) operators, $\mathbf{S}_i = C_i^{\dagger}\sigma C_i/2$ are spin operators with $\sigma = (\sigma_x, \sigma_y, \sigma_z)$ as Pauli matrices, $\mu$ is the chemical potential, and the summation $\langle ij \rangle$ is carried over nearest bonds. In this paper, we study the charge dynamics of copper oxide materials in the underdoped regime within the $t$-$J$ model.

The strong electron correlation in the $t$-$J$ model manifests itself by the electron single occupancy on-site local constraint. This is why the crucial requirement is to impose this electron on-site local constraint for a proper understanding of the physics of copper oxide materials. At the half-filling, the $t$-$J$ model is reduced as the Heisenberg model. In this case, it has been shown that the physical properties of the Heisenberg model can be well described within the Schwinger boson theory[4]. Then for the $t$-$J$ model, a nature choice is the slave-fermion Schwinger boson theory[5], where the constrained electron operators $C_{i\sigma}$ is mapped onto the slave-fermion Schwinger boson formalism as $C_{i\sigma} = h_i^{\dagger} a_{i\sigma}$ with the spinless fermion operator $h_i$ keeps track of the charge (holon), while the boson operators $a_{i\sigma}$ keeps track of the spin (spinon), and the $t$-$J$ model (1) can be expressed in the slave-fermion Schwinger boson representation as,

$$H = -t \sum_{\langle ij \rangle \sigma} h_j h_i^{\dagger} a_{i\sigma}^{\dagger} a_{j\sigma} + h.c. - \mu \sum_i h_i^{\dagger} h_i + J \sum_{\langle ij \rangle} \mathbf{S}_i \cdot \mathbf{S}_j, \quad (2)$$

supplemented by the on-site local constraint $h_i^{\dagger} h_i + \sum_{\sigma} a_{i\sigma}^{\dagger} a_{i\sigma} = 1$, where the spin operators $\mathbf{S}_i = a_i^{\dagger} \sigma a_i/2$. This on-site local constraint restricts both holons and spinons, and always replaced by a global one in analytical calculations since this local constraint is very cumbersome to handle exactly[5]. To avoid this weakness, we solve this constraint by rewritten the boson operators $a_{i\sigma}$ in terms of the CP$^1$ boson $b_{i\sigma}$ as $a_{i\sigma} = (1 - h_i^{\dagger} h_i)^{1/2} b_{i\sigma}$. In this case, the $t$-$J$ model (2) can be rewritten in the slave-fermion CP$^1$ boson representation as,

$$H = -t \sum_{\langle ij \rangle \sigma} h_j h_i^{\dagger} b_{i\sigma}^{\dagger} b_{j\sigma} + h.c. - \mu \sum_i h_i^{\dagger} h_i + J \sum_{\langle ij \rangle} (h_i h_i^{\dagger}) \mathbf{S}_i \cdot \mathbf{S}_j (h_j h_j^{\dagger}), \quad (3)$$

supplemented by the on-site local constraint $\sum_{\sigma} b_{i\sigma}^{\dagger} b_{i\sigma} = 1$, where the pseudospin operators $\mathbf{S}_i = b_i^{\dagger} \sigma b_i/2$. Now only spinons are restricted by the on-site local constraint, but the effect of the holon on the spinon in the on-site local constraint in the slave-fermion Schwinger boson formalism has been reflected by the factor $(1 - h_i^{\dagger} h_i)^{1/2}$ in the slave-fermion CP$^1$ formalism. In this case, it has been shown within the fermion-spin theory[6] that the constrained CP$^1$ bosons $b_{i\sigma}$ can be mapped exactly onto the pseudospin representation as $b_{i\uparrow} = S_i^-$ and $b_{i\downarrow} = S_i^+$, where the on-site local constraint $\sum_{\sigma} b_{i\sigma}^{\dagger} b_{i\sigma} = S_i^+ S_i^- + S_i^- S_i^+ = 1$ is exactly satisfied. Therefore in the present fermion-spin formalism, the spinless fermion operator $h_i$ describes the charge (holon) degrees of freedom, while the operators $(1 - h_i^{\dagger} h_i)^{1/2} S_i$ together describe the spin (spinon) degrees of freedom, and the low-energy behavior of the $t$-$J$ model (3) can be written in the fermion-spin representation as [6],

$$H = -t \sum_{\langle ij \rangle \sigma} h_j h_i^{\dagger}(S_i^+ S_j^- + S_i^- S_j^+) + h.c. - \mu \sum_i h_i^{\dagger} h_i + J_{eff} \sum_{\langle ij \rangle} \mathbf{S}_i \cdot \mathbf{S}_j, \quad (4)$$

where $J_{eff} = [(1-\delta)^2 - \phi^2]$, $\delta$ is the hole doping concentration, $\phi = \langle h_i^{\dagger} h_{i+\hat{\eta}} \rangle$ is the holon particle-hole order parameter, $\hat{\eta} = \pm \hat{x}, \pm \hat{y}$.

Within the decoupling scheme, the in-plane spinon and holon Green's functions are defined as $D_{ab}(i-j, \tau - \tau') = -\langle T_{\tau}[1 - h_i^{\dagger}(\tau) h_i(\tau)]^{1/2} S_i^+(\tau) S_j^-(\tau')[1 - h_j^{\dagger}(\tau') h_j(\tau')]^{1/2}\rangle$ and $g_{ab}(i-j, \tau - \tau') = -\langle T_{\tau} h_i(\tau) h_j^{\dagger}(\tau') \rangle$, respectively. In the underdoped regime ($\delta \ll 1$), this in-plane spinon Green's function can be expressed approximately as $D_{ab}(i-j, \tau - \tau') \approx -(1-\delta)\langle T_{\tau} S_i^+(\tau) S_j^-(\tau') \rangle \approx -\langle T_{\tau} S_i^+(\tau) S_j^-(\tau')\rangle$. In this case, the in-plane conductivity has been obtained based on the Ioffe-Larkin combination rule as [7],

$$\sigma_{ab}(\omega) = (4te\chi)^2 \frac{2}{N} \sum_k \gamma_{sk}^2 \int_{-\infty}^{\infty} \frac{d\omega'}{2\pi} A_h(k, \omega') A_h(k, \omega' + \omega) \frac{F(\omega, \omega')}{\omega}, \quad (5)$$

where $\gamma_{sk} = (\sin k_x + \sin k_y)/2$, $n_F(\omega)$ and $n_B(\omega)$ are the Fermi and Bose distribution functions, respectively, $\chi = \langle S_i^+ S_{i+\hat{\eta}}^- \rangle$ is the spinon bond amplitude, $F(\omega, \omega') = n_F(\omega' + \omega) - n_F(\omega')$, and $A_h(k, \omega) = -2\mathrm{Im} g_{ab}(k, \omega)$ is the in-plane holon spectral function obtained by considering the fluctuation to the second-order. In the underdoped regime, the experimental results show that the ratio $R = \rho_c(T)/\rho_{ab}(T)$ ranges from $R \sim 100$ to $R > 10^5$, this large magnitude of the resistivity anisotropy reflects that the c-axis mean free path is shorter than the interlayer distance, and the carriers are tightly confined to the $CuO_2$ planes, and also is the evidence of the incoherent charge dynamics in the c-axis direction [1,2]. For the chain copper oxide materials, the presence of the rather conductive Cu-O chains in the underdoped regime can reduce the blocking effect [1,2], therefore the c-axis charge dynamics in this system is effected by the same electron interaction as that in the in-plane, and the c-axis conductivity can be calculated in terms of the in-plane holon spectral function by using standard formalisms for the tunneling in metal-insulator-metal junctions as [8],

$$\sigma_c(\omega) = \frac{1}{2}(4\tilde{t}_c e \chi c_0)^2 \frac{1}{N} \sum_k \int_{-\infty}^{\infty} \frac{d\omega'}{2\pi} A_h(k, \omega') A_h(k, \omega' + \omega) \frac{F(\omega, \omega')}{\omega}, \quad (6)$$

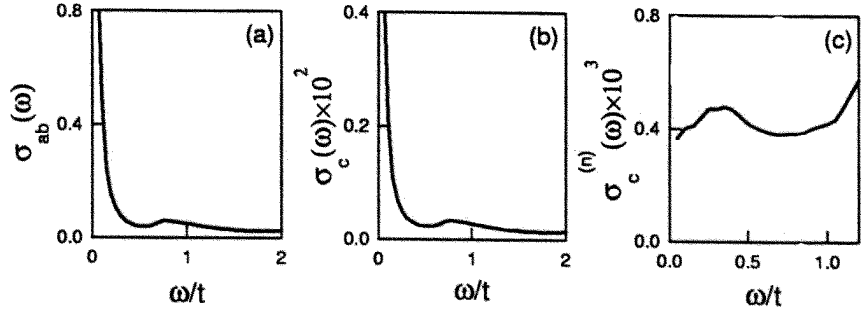

Figure 1: (a) $\sigma_{ab}(\omega)$, (b) $\sigma_c(\omega)$, and (c) $\sigma_c^{(n)}(\omega)$ at the doping $\delta = 0.06$ for the parameters $t/J = 2.5$, $\bar{t}_c/t = 0.04$ with $T = 0$.

where $c_0$ is the interlayer distance, $\tilde{t}_c$ is the weak interlayer hopping matrix element. For the no-chain copper oxide materials, the doped holes may introduce a disorder in between the $CuO_2$ planes. Moreover, the disorder introduced by doped holes residing between the $CuO_2$ planes modifies the interlayer hopping elements as the random matrix elements. In this case, the corresponding momentum-nonconserving expression of the c-axis conductivity $\sigma_c^{(n)}(\omega)$ is obtained by the replacement of the in-plane holon spectral function $A_h(k,\omega)$ in Eq. (6) with the in-plane holon density of states (DOS) $\Omega_h(\omega) = 1/N \sum_k A_h(k,\omega)$ as [9],

$$\sigma_c^{(n)}(\omega) = \frac{1}{2}(4\bar{t}_c e \chi c_0)^2 \int_{-\infty}^{\infty} \frac{d\omega'}{2\pi} \Omega_h(\omega' + \omega) \Omega_h(\omega') \frac{F(\omega,\omega')}{\omega}, \qquad (7)$$

where $\bar{t}_c$ is some average of the random interlayer hopping matrix elements. We have performed a numerical calculation for the conductivity, and the results of (a) $\sigma_{ab}(\omega)$, (b) $\sigma_c(\omega)$, and (c) $\sigma_c^{(n)}(\omega)$ at the doping $\delta = 0.06$ for the parameters $t/J = 2.5$, $\bar{t}_c/t = 0.04$ with the temperature $T = 0$ is plotted in Fig. 1, where $\sigma_{ab}(\omega)$ and $\sigma_c(\omega)$ show that there are a low-energy peak at $\omega < 0.5t$ separated by a gap or pseudogap $\approx 0.4t$ from the broad absorption band (midinfrared band) in the conductivity spectrum, the midinfrared peak is doping dependent and the peak is shifted to lower energy with increased doping; while $\sigma_c^{(n)}(\omega)$ contains two bands, the higher-energy band shows a broad peak at $\sim 0.3t$. The weight of this band is increased with dopings, but the peak position does not appreciably shift to lower energies. As a consequence of this pinning of the transferred spectral weight, the weight of the lower-energy band, corresponding to the non-Drude peak in $\sigma_{ab}(\omega)$ and $\sigma_c(\omega)$, is quite small and does not form a well-defined peak at low energies. These results are qualitatively consistent with the experimental results of the underdoped copper oxide materials[1,2].

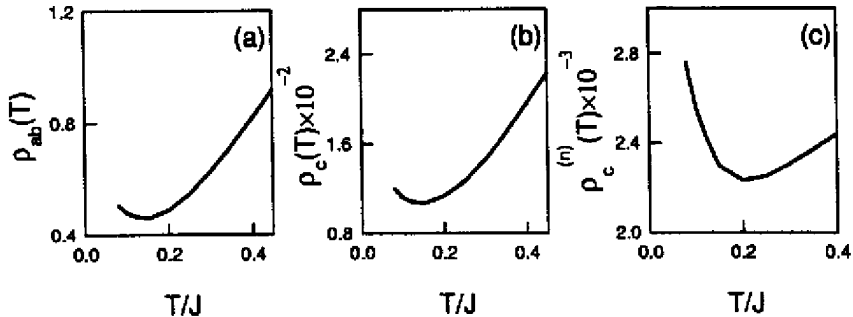

Figure 2: (a) $\rho_{ab}$, (b) $\rho_c$, and (c) $\rho_c^{(n)}$ at the doping $\delta = 0.06$ for $t/J = 2.5$, $\bar{t}_c/t = 0.04$, and $c_0/a_0 = 2.5$.

For the further understanding the behavior of the charge dynamics of the underdoped copper oxide materials, we have also performed the numerical calculation for the resistivity $\rho_\alpha = 1/\lim_{\omega \to 0} \sigma_\alpha(\omega)$, and the results of (a) $\rho_{ab}(T)$, (b) $\rho_c(T)$, and $\rho_c^{(n)}(T)$ at the doping $\delta = 0.06$ for the parameters $t/J = 2.5$, $\bar{t}_c/t = 0.04$, and $c_0 = 2.5a_0$ are plotted in Fig. 2. These results show that $\rho_{ab}(T)$ exhibits a temperature linear dependence with deviations at low temperatures, while $\rho_c(T)$ and $\rho_c^{(n)}(T)$ are characterized by a crossover from the high temperature metallic-like to the low temperature semiconducting-like. Moreover, the values of $\rho_c(T)$ and $\rho_c^{(n)}(T)$ are by $2 \sim 4$ orders of magnitude larger than these of $\rho_{ab}(T)$ in the corresponding energy range, which are also qualitatively consistent with the experimental results of the underdoped copper oxide materials [1,2]. Our results also show that the crossover to the semiconducting-like range in $\rho_c(T)$ is obviously linked with the crossover from the temperature linear to the nonlinear range in $\rho_{ab}(T)$.

It has been shown that an remarkable point of the pseudogap is that it appears in both of spinon and holon excitations. The present study indicates that the observed crossovers of $\rho_{ab}$ and $\rho_c$ seem to be connected with the holon pseudogap, which can be understood from the physical property of the in-plane holon DOS $\Omega_h(\omega)$. The numerical result of $\Omega_h(\omega)$ at the doping $\delta = 0.06$ as a function of energy for the temperature (a) $T = 0$ and (b) $T = 0.2J$ is plotted in Fig. 3, which shows that there is a V-shape holon pseudogap near the chemical potential $\mu$. This holon pseudogap decreases with increasing temperatures, and vanishes at higher temperatures. Moreover, we have found that the holon pseudogap grows monotonously as the doping $\delta$ decreases, and disappear in the overdoped regime. This holon pseudogap would reduce the in-plane holon scattering and thus is responsible for the metallic to semiconducting crossover in the c-axis resistivity $\rho_c$ and the deviation from the temperature linear behavior in the in-plane resistivity $\rho_{ab}$.

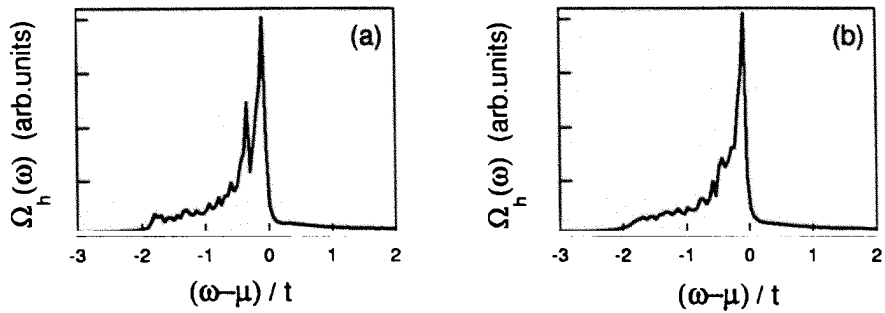

Figure 3: $\Omega_h(\omega)$ at the doping $\delta = 0.06$ for $t/J = 2.5$ with (a) $T = 0$, and (b) $T = 0.2J$.

In summary, we have studied the charge dynamics of the underdoped copper oxide materials within the $t$-$J$ model. Our results show that both in-plane charge dynamics and c-axis charge dynamics are mainly governed by the scattering from the in-plane fluctuation, which would be suppressed when the holon pseudogap opens at low temperatures, leading to the temperature linear to the nonlinear range in the in-plane resistivity and crossovers to the semiconducting-like range in the c-axis resistivity.